\documentclass[aps,prapplied,reprint,twocolumn]{revtex4-2}
\usepackage{amssymb}
\usepackage{graphicx}
\usepackage{amsmath}
\usepackage{epsfig}
\usepackage[latin9]{inputenc}
\usepackage{array}
\usepackage{multirow}
\usepackage{color}
\usepackage{esint}
\usepackage{bm}
\usepackage{bbm}
\usepackage{epstopdf}
\usepackage{mathtools}
\usepackage{soul}
\usepackage[dvipsnames]{xcolor}
\usepackage{tikz}
\usepackage{tabularx}
\usepackage{float}
\usepackage[bookmarks=true,pdftex,colorlinks=true,urlcolor=blue,linkcolor=black,citecolor=blue,breaklinks=true,hypertexnames=false]{hyperref}
\usepackage{multibib} 
\usepackage{physics}
\usepackage{makecell}
\usepackage{booktabs}
\begin{document}
\title{Two-color magneto-optical trapping of ytterbium atoms}

\author{Xiao Li}
\affiliation{State Key Laboratory for Mesoscopic Physics and Frontiers Science Center for Nano-optoelectronics, School of Physics, Peking University, Beijing 100871, China}

\author{Yufei Wang}
\affiliation{State Key Laboratory for Mesoscopic Physics and Frontiers Science Center for Nano-optoelectronics, School of Physics, Peking University, Beijing 100871, China}

\author{Ligeng Yu}
\affiliation{State Key Laboratory for Mesoscopic Physics and Frontiers Science Center for Nano-optoelectronics, School of Physics, Peking University, Beijing 100871, China}

\author{Bo Song}\email{bsong@pku.edu.cn}
\affiliation{State Key Laboratory for Mesoscopic Physics and Frontiers Science Center for Nano-optoelectronics, School of Physics, Peking University, Beijing 100871, China}

\begin{abstract}

We report laser cooling and trapping of ytterbium atoms in a two-color magneto-optical trap (MOT). Benefited from both the broad singlet transition ($^1\text{S}_0\rightarrow {}^1\text{P}_1$) and the narrow intercombination transition ($^1\text{S}_0\rightarrow {}^3\text{P}_1$) of ytterbium atoms, the two-color MOT enables rapid loading and efficient cooling. We systematically investigate the shielding effect of the intercombination transition by examining the atom loading and loss rates of single-color and two-color MOTs. Our findings are general and can be extended to other alkaline earth(-like) atoms.
 
\end{abstract}
    
\maketitle

\section{Introduction}
Alkaline-earth or alkaline-earth-like atoms such as strontium and ytterbium, featuring two-valence-electron structures and (ultra)narrow line transitions, enable numerous applications ranging from quantum computing and simulation, to atom interferometers and precision measurements~\cite{finkelstein2024universal,reichardt2024logical,aeppli2024clock,hu2017atom,schafer2020tools}. Fast laser cooling and trapping of cold atoms are essential, including Doppler cooling, Zeeman slowing, and magneto-optical trapping~\cite{metcalf1999laser,phillips1998nobel,ketterle2002nobel,schreck2021laser}. Recently, magneto-optical trapping of alkaline-earth(-like) atoms was demonstrated using both the broad singlet transition ($^1\text{S}_0\rightarrow {}^1\text{P}_1$) and the narrow intercombination transition ($^1\text{S}_0\rightarrow {}^3\text{P}_1$) ~\cite{kawasaki2015two,seo2020efficient,lunden2020enhancing,plotkin2020crossed,bondza2022two,bloch2024anisotropic,wang2025two,lee2015core}. 
However, a systematic understanding of magneto-optical trap (MOT) optimization remains challenging, particularly in the presence of interactions and different transitions, such as those involved in losses and shielding enhancement~\cite{hoschele2023atom, wang2025two}. 
A deeper understanding and optimized control of the MOT, the loading process and loss mechanisms in particular, could enable rapid and robust preparation of large atom numbers for quantum computing and simulation with neutral atoms.

Here, we report an experimental setup consisting of a Zeeman slower and a two-color MOT to achieve fast loading and cooling of $^{174}\text{Yb}$ atoms with large steady-state numbers. We investigated the atom loading process in three configurations, blue (the $6\text{s}^2\,{}^1\text{S}_0\rightarrow 6\text{s}6\text{p}\,{}^1\text{P}_1$ transition at $399\,$nm), green (the $6\text{s}^2\,{}^1\text{S}_0\rightarrow 6\text{s}6\text{p}\,{}^3\text{P}_1$ transition at $556\,$nm) and two-color MOTs. By comparing both the loading and loss rates of them, we systematically studied the interplay of singlet and intercombination transitions, providing further insight into the role of shielding enhancement in suppressing losses and boosting the MOT number.  

\section{Experimental setup}
\begin{figure}[htbp]
    \centering
    \includegraphics[width=1\linewidth]{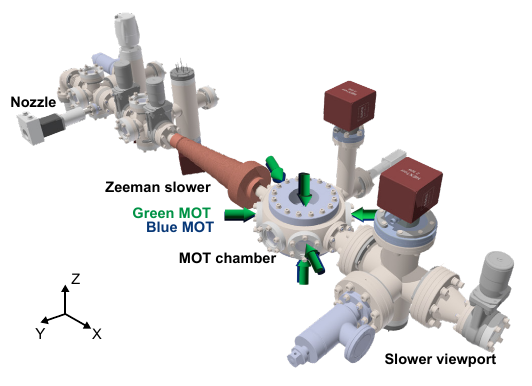}
    \caption{\textbf{Experimental setup}. Schematic of the experimental apparatus includes (from top left to bottom right) an oven chamber, a Zeeman slower, a magneto-optical trap (MOT) and a glass cell. The two-color MOT consists of three pairs of blue and green beams (indicated by arrows) and a pair of anti-Helmholtz coils. The slower light is shined onto the atomic flux through the bottom right viewport.
\label{fig:experiment setup}}
\end{figure}

Our apparatus consists of an atomic oven, two six-way crosses, a Zeeman slower, a MOT chamber, and a glass cell in Fig.~\ref{fig:experiment setup}. The first two crosses are designed for spectroscopy and transverse cooling. The atomic nozzle between the oven and the first six-way cross, composed of multiple individual nozzles arranged in a hexagonal geometry with an aspect ratio of 1:50, efficiently collimates the atomic flux in the transverse direction while recycling atoms back to the oven. 
Together with a differential tube between the Zeeman slower and the second six-way cross, the Seas NEXTorr Z200 and Z300 UHV pumps can maintain the vacuum pressure of the MOT chamber on the order of $10^{-11}\,$mbar. The Zeeman slower generates a position-dependent magnetic field and compensates for the Doppler shift of the slowed atoms, facilitating cooling along the $X$ direction. The MOT consists of an octagon chamber, five CF35 viewports, and two CF100 recessed viewports for good optical access that are anti-reflection coated for wavelengths of both $399\,$nm and $556\,$nm. Three pairs of blue and green MOT beams are shined into the chamber indicated by arrows in Fig.~\ref{fig:experiment setup}. 
The magnetic field of the MOT is a pair of anti-Helmholtz coils, together with three pairs of bias coils to compensate for the residual magnetic field, generating the designed magnetic field.

\section{Zeeman slower}
The Zeeman slower is designed to compensate the Doppler shift during laser cooling of atoms.
The slower beam is shined through the viewport and counter-propagates to the atomic flux in Fig.~\ref{fig:experiment setup}. The oven temperature is 390$\,^\circ$C and the corresponding peak velocity is $305\,$m/s. The capture velocity of the MOT along the $X$ direction is designed to be $\sim10 \,$m/s. 
The estimated flux of $^{174}$Yb atoms reaching the MOT region is $2\times10^{11}\,$s$^{-1}$.

The total length of the Zeeman slower is $32.5\,$cm, and it is wound using two types of copper coils along a 1-inch diameter 316LN stainless steel tube. The main coil consists of a diameter of $1.6\,$mm and is supplied by a small current $I_s\sim3.9\,$A. The other used to boost the magnetic field, the so-called boost coil, is made of a square hollow copper tube with an outer width of 3$\,$mm and an inner diameter of 1$\,$mm. The hollow shape of the coil is for water cooling. To achieve the designed magnetic-field profile, each layer is wounded using single-pitch and double-pitch configurations~\cite{song2016cost}, with details in Table~\ref{tab:coil}. 
The power of the slower light is about $20\,$mW and the frequency is $756\, $MHz red-detuned from the $^1\text{S}_0\rightarrow{}^1\text{P}_1$ transition. The slower light is focused down from 8 mm at the MOT center to $5\,$mm in diameter at the oven nozzle to match the atomic beam divergence.
\begin{figure}[htbp]
    \centering
    \includegraphics[width=1\linewidth]{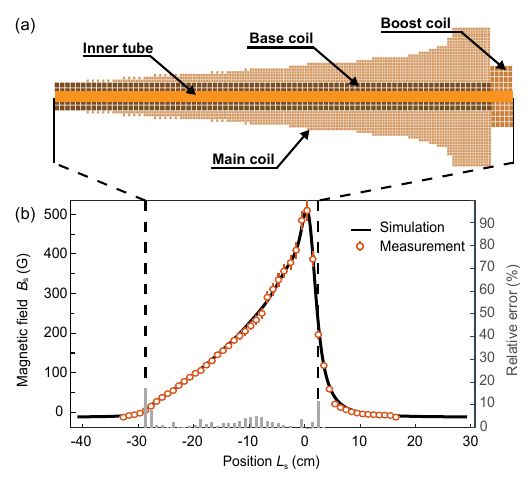}
    \caption{\textbf{Zeeman slower.} (a) The Zeeman slower consists of an inner tube, water-cooled base coils, main coils and water-cooled boost coils. (b) The measured magnetic field (open circles) as a function of position agrees well with the simulation (solid line). Here, the test currents of the main and the boost coils are set to $3.6\,$A and $43\,$A, respectively. Errorbars represent the uncertainty of the measurement. The Zeeman slower is designed for slowing atoms at a temperature of $390\,^\circ$C, with the slower light at $\sim20\,$mW and $756\,$MHz red-detuned from the $^1\text{S}_0\rightarrow {}^1\text{P}_1$ transition. 
\label{fig:bfield Schematic}}
\end{figure} 
\begin{table}[htbp]
\centering
\begin{tabular}{lcccccc}
\toprule[1pt]
 Layer  & \makecell[c]{Single- \\pitch}  &\makecell[c]{Double- \\pitch}& Layer &\makecell[c]{Single- \\pitch} &\makecell[c]{Double- \\pitch} \\
   \midrule[1pt]
  $\overline{n_{1,2}} $ & 83  & 0 & $\textit{n}_9$  & 60  & 0 &         \\
 $\textit{n}_1$  & 171  & 5  &   $\textit{n}_{10}$  & 45  & 0           \\
 $\textit{n}_2$  & 158  & 4 &   $\textit{n}_{11}$  &25 &0            \\
 $\textit{n}_3$  & 140  & 3 &  $\textit{n}_{12}$   &24  &0            \\
 $\textit{n}_4$  & 126  & 3 &  $\textit{n}_{13-16}$   &23  &0             \\
 $\textit{n}_5$  & 108  & 1 &  $\textit{n}_{17-19}$   &22  &0              \\
 $\textit{n}_6$  & 95  & 0 &  $\textit{n}_{20}$   &21  &0            \\
 $\textit{n}_7$  & 82  & 0 &   Boost coil      & 5 $\times$ 5              \\
 $\textit{n}_8$  & 67  & 0 &               \\
\bottomrule[1pt]
\end{tabular}
\caption{\textbf{Coil details of the Zeeman slower.} The coils are wound using two types of spacing: single pitch and double pitch. In the single-pitch configuration, the coils are tightly wound, while for the double pitch, they are spaced apart by a distance equal to coil's thickness. $\overline{n_i}$ represents the number of turns for the base coils, which are water cooled and designed to provide an offset magnetic field. Here, $n_i$ denotes the number of turns at the $i_{th}$ layer starting from the slower exit position for the base and main slower coil, respectively. The double-pitch coils are added for the first five layers after the single-pitch coils. The boost coils are constructed by $5\times5$ water-cooled coils.}
\label{tab:coil}
\end{table}

\section{Shielding enhancement}
An essential requirement for laser cooling and trapping is a closed cycling transition of atoms. Although the blue MOT using the singlet transition $^1\text{S}_0\rightarrow {}^1\text{P}_1$ at $399\,$nm has a high loading rate, the ${}^1\text{P}_1$ state suffers from leakage. As a result, a branch ratio of approximately $7\times 10^{-8}$ in the ${}^1\text{P}_1$ state decay to the $^3$D$_j (j=1,2)$ with rates of $2.2 \,\text{Hz}$ and $9.6\,\text{Hz}$, followed by decaying to the $^3$P$_j (j=0,1,2)$ states, respectively~\cite{honda1999magneto,cho2012optical,loftus2000power,letellier2023loading}.
Notably, atoms in the $^3\text{P}_j$ state are not trapped in the MOT and escape from the trap~\cite{honda1999magneto}. Partial atoms in the intercombination state $^3\text{P}_1$ can spontaneously decay to the $^1\text{S}_0$ state with rates of $1.1\,\text{MHz}$, whereas the $^3\text{P}_0$ and $^3\text{P}_2$ states are metastable with rates less than $0.1\,$Hz.
Our method to reduce the atom loss in the blue MOT is to implement additional green light (the $^1\text{S}_0\rightarrow {}^3\text{P}_1$ transition). Since the singlet transition ($^1\text{S}_0\rightarrow {}^1\text{P}_1$) is unclosed and $^3$P$_j (j=0,1,2)$ states are not trapped, resulting in the atom loss. By strongly coupling the $^3\text{P}_1$ and $^1\text{S}_0$ states, atoms in the $^3\text{P}_1$ state can be effectively trapped by the blue MOT without altering the population balance between the $^1\text{S}_0$ and $^1\text{P}_1$ states. Therefore, the $^3\text{P}_1$ state can act as a reservoir and shield trapped atoms from the loss of unclosed channels~\cite{hoschele2023atom, wang2025two}. 

We explore this shielding enhancement in the two-color MOT under experimentally optimized conditions. Both one-body ($\alpha N$) and two-body losses ($\beta N^2$) are taken into account, and the rate equation describing loading and holding atoms in the MOT is expressed as
\begin{equation}
    \frac{\dd N(t)}{\dd t} = \eta -\alpha N(t)-\beta N(t)^2 \label{rate equation},
\end{equation}
where $N(t)$ is the atom number of the MOT, $\eta$ is the loading rate, $\alpha$ and $\beta$ are coefficients of one-body and two-body losses. The final steady-state atom number can be approximated as
$N = (\sqrt{\alpha^2+4\beta\eta}-\alpha)/2\beta$. We analyze the enhancement of shielding by considering a three-level energy diagram of Yb atoms, with populations $N_\text{s}$, $N_\text{b}$ and $N_\text{g}$ in the ground state $^1\text{S}_0$, excited state $^1\text{P}_1$ and intercombination state $^3\text{P}_1$ respectively. The one-body loss rate $\alpha$ is attributed by three sources, background gas collisions $\gamma_{\text{c}}$, the loss from the $^1\text{P}_1$ state $\gamma_{\text{b}}$, and the loss from the $^3\text{P}_1$ state $\gamma_{\text{g}}$, which is negligible~\cite{cho2012optical}. The one-body loss rate of the blue MOT can be expressed as $\alpha_\text{b}=\gamma_c+\gamma_{\text{b}}N_\text{b}/(N_\text{b}+N_\text{s})$, while for the green MOT, $\alpha_\text{g} = \gamma_c+\gamma_{\text{g}}N_{\text{g}}/(N_{\text{g}}+N_\text{s})$ and for the two-color MOT, $\alpha_{2} = \gamma_c+(\gamma_{\text{b}}N_{\text{b}}+\gamma_{\text{g}}N_{\text{g}})/(N_\text{b}+N_\text{g}+N_\text{s})$. In the blue MOT only, the population of the $^3\text{P}_1$ state can be neglected since the transition from decaying channels to the $^3\text{P}_{1}$ state is much slower than the transition from the $^3\text{P}_{1}$ to the $^1\text{S}_{0}$ states.

The population ratio between the upper and lower states is generally given by $s/(2+s)$, where the coefficient $s = (I/I_\text{sat})/(1+(2\delta/\Gamma)^2)$ depends on the laser detuning $\delta$ and intensity of the laser $I$, the saturation intensity $I_\text{sat} = (\pi h c \Gamma)/(3\lambda^3)$ with Planck's constant $h$, the speed of light $c$ and the wavelength of light $\lambda$.  

\begin{figure}[htbp]
    \centering
    \includegraphics[width=1\linewidth]{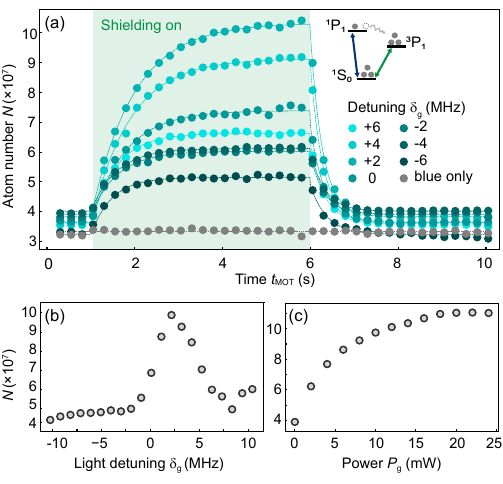}
    \caption{\textbf{Shielding enhancement.} (a) Atom loading in the blue MOT ($^1\text{S}_0\rightarrow {}^1\text{P}_1$) is enhanced by two pairs of horizontal green beams ($^1\text{S}_0\rightarrow {}^3\text{P}_1$). After the blue MOT reaches saturation, the green light is turned on from $t_\text{MOT}=1$ to $6\,$s. In contrast to the steady-state MOT number, the initial loading rate is less sensitive to the detuning of the green light. The power per beam is $22.5\,$mW. The dashed lines represent fits to the data using exponential growth and decay curves. (b) The saturated MOT number is probed as a function of the detuning of the green light. (c) The enhancement of the green light is saturated at around 20$\,$mW per beam. Here the detuning is set to $+2\,$MHz.
\label{fig:enhancement}}
\end{figure}

Two pairs of green beams are applied in the horizontal plane to the blue MOT and probed the loading process. To avoid a three-dimensional green MOT and study the shielding effect only, the vertical pair of green beams is absent.  Figure~\ref{fig:enhancement}(a) shows a clear shielding enhancement in both the loading rate and the steady-state number. The MOT number is probed by absorption imaging using the singlet transition at $399\,$nm. After the blue MOT loading reaches saturation, the green light is turned on at $t_1$ and off at $t_2$. The dashed lines represent fits to the data using $N(t)=N_\text{b} +\Delta N(1-\exp(-(t-t_1)/\tau_1))$, while $N(t) =N_1 +\Delta N\exp(-(t-t_2)/\tau_2)$ as growth and decay curves, serving as guides for eyes, where $N_\text{b}$ and $\Delta N$ are the blue MOT number and increase due to the shielding light. The max loading enhancement occurs at $\delta_\text{g} \simeq 2\pi \times 2\,\text{MHz}=11\,\Gamma_\text{g}$ blue detuned from the resonance (also see Fig.~\ref{fig:enhancement}(b)), which agrees with findings in Ref.~\cite{wang2025two}. Moreover, the initial loading rates are dominated by the blue MOT and thus are similar for different detuning.

Figure~\ref{fig:enhancement}(c) shows the atom number as a function of the green light power, where the interplay of photon scattering and the shielding effect gives rise to a saturated value. the shielding effect saturates as the power reaches approximately $20\,$mW, constrained by the population distribution following $s/(2+s)$.

\section{Two-color MOT}
Using the benefits of both transitions, the two-color MOT can provide a fast loading rate while maintaining a large steady-state atom number.  The MOT power per beam in the direction $X$, $Y$, $Z$ is $(15.5,15.5,8)\,$mW and corresponding intensities are $(0.33,0.33,0.17)\,I_\text{sat}^\text{b}$. The power of the green MOT is $(30.5,30.5,13)\,$mW and corresponding intensities of $(278,278,118)\,I_\text{sat}^\text{g}$. Here, the saturation intensities for the blue and green MOTs are $I_\text{sat}^\text{b}=59.9\,\text{mW}/\text{cm}^2$ and $I_\text{sat}^\text{g}=0.14\,\text{mW}/\text{cm}^2$, respectively. The detuning of the blue and green MOTs are $\delta_{\text{b}}=-1.2\Gamma_{\text{b}}$, and $\delta_{\text{g}}=-33\Gamma_{\text{g}}$, respectively. The optimized magnetic-field gradients for blue and green MOTs are $\partial B_\text{b}/\partial z =2\partial B_\text{b}/\partial x=2\partial B_\text{b}/\partial y=24\,$G/cm and $\partial B_\text{g}/\partial z$= $4.2\,$G/cm. The optimized parameters are $\partial B_\text{2}/\partial z$= $29.4\,$G/cm for the two-color MOT and $\partial B_\text{2}/\partial z$= $1.4\,$G/cm for the final green MOT, which are consistent with findings in Ref.~\cite{lee2015core}. Note that the two-color MOT not only has a rapid loading rate, but is found to be robust against MOT power imbalance even at low magnetic field gradients~\cite{kawasaki2015two}.
Two-color MOT has the same power of the blue beams and detuning as the blue MOT $\delta_\text{b} = -1.2\Gamma_\text{b}$. We optimize the shielding effect in the two-color MOT by varying $\delta_\text{g}$ from $+33\Gamma_\text{g}$ to $-33\Gamma_\text{g}$, and set the frequency of the green beams to $\delta_\text{g}=+2\pi\times2\,$MHz.

\begin{figure}[t]
    \centering
    \includegraphics[width=1\linewidth]{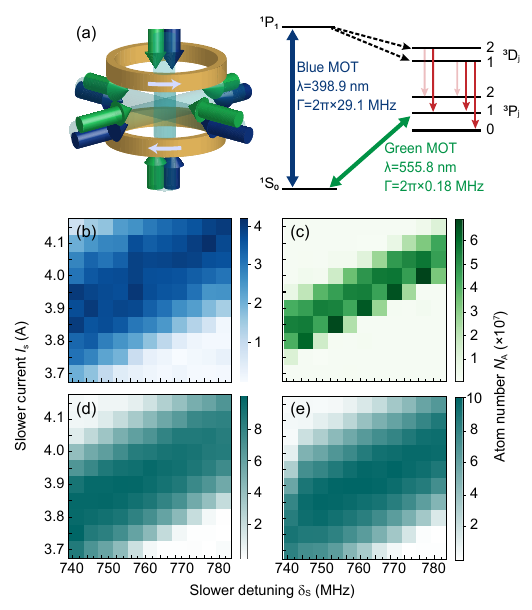}
    \caption{\textbf{MOT loading under different transitions and configurations.} (a) The two-color MOT setup and the energy level diagram of $^{174}$Yb atoms. MOT beams indicated by arrows are aligned together and displaced for clarity. The singlet transition $^1\text{S}_0\rightarrow {}^1\text{P}_1$ is not closed. Atoms in the ${}^1\text{P}_1$ state can decay to $^3\text{D}_j (j=1,2)$ states and further to $^3\text{P}_j (j=0,1,2)$ states, before either escaping or returning to the $^1\text{S}_0$ state. (b,c) are MOT loading of the blue MOT and green MOT probed as a function of the slower current $I_{\text{s}}$ and detuning $\delta_{\text{s}}$ from the resonant singlet transition. The power of the Zeeman slower beam is $23\,$mW. (d,e) are the two-color MOT with the power of the slower light of $18\,$mW and $23\,$mW, respectively. The MOT load time is $5\,$s in (b,d,e) and $80\,$s (c).
\label{fig:detuningvscurrent}}
\end{figure}

Figure~\ref{fig:detuningvscurrent} shows the loading of the single-color and two-color MOTs as a function of the current of the Zeeman slower and the detuning. The increase in the initial slower detuning $\delta_s$ should be compensated by an increase in the magnetic field of the slower. Around the optimal regime, the MOT atom number remains nearly constant as long as $(I_\text{s},\delta_\text{s})$ lies within the optimized linear regime. Compared to the blue MOT, the green MOT exhibits a narrower distribution because of its narrower linewidth, while the width of the two-color MOT is mainly dominated by the blue MOT. 

The capture velocity of the blue MOT is estimated as $v_{\text{b}}=\sqrt{\hbar k_\text{b} \Gamma_\text{b}d /m}= 103\,$m/s where the natural linewidth is $\Gamma_\text{b}$, the MOT beam size $d=10\,\text{mm}$, $\hbar$ the reduced Planck constant, $m$ the atomic mass and $k_\text{b}=2\pi/\lambda_\text{b}$. The blue MOT can capture atoms with final velocities slower than $103\,$m/s, and cool them down to several times the Doppler limit $T_\text{D}^\text{b}=\hbar \Gamma_\text{b}/2k_\text{B}=698\,\mu$K, $\sim7\,$mK in our experiment, where $k_\text{B}$ is the Boltzmann constant.
However, for the green MOT, the capture velocity is $v_\text{g}=7\,$m/s and the Doppler limit is $T_\text{D}^\text{g}= 4.4\,\mu$K. 
Correspondingly, the atoms can be further cooled down to $\sim 20\,\mu$K in the green MOT in our experiment, comparable to Ref.~\cite{lee2015core}.

We further optimize the atom transfer from the two-color MOT to a final pure green MOT. The blue MOT is switched off within 1 ms, and the green MOT frequency is suddenly changed to $33\, \Gamma_\text{g}$ red-detuned from the resonance, and the magnetic field is changed from  $\partial B_\text{b}/\partial z$ to $\partial B_\text{g}/\partial z$. Finally, the MOT was compressed for loading into an optical dipole trap and further cooling. The detuning of the green MOT was quickly shifted from $\delta_\text{g}=-33\Gamma_\text{g}$ to $\delta_\text{g}\simeq -8\Gamma_\text{g}$, and the power of the green MOT was ramped down from $30.5\,$mW to $35\,\mu$W in $100\,$ms. Simultaneously, the magnetic field gradient is changed to $\partial B/\partial z=18.2\,$G/cm. The MOT transfer and compression are efficient with a loss rate of $\sim 10\%$, and provides a compressed MOT of $10^8\,$ atoms in a diameter of around 300$\,\mu$m  with a temperature of 18$\,\mu$K.

\begin{figure}[t]
    \centering
    \includegraphics[width=1\linewidth]{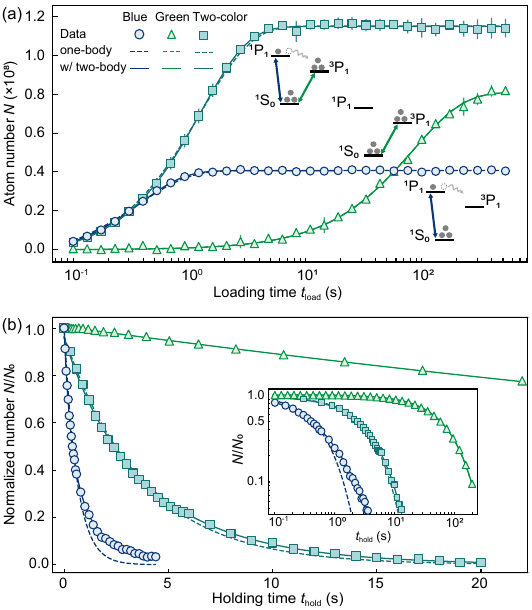}
    \caption{\textbf{MOT loading and holding curves}. (a) Atom numbers are measured as a function of loading time for different MOTs. The two-color MOT (cyan squares) benefited from rapid loading of the blue MOT (blue circles) and further cooling of the green MOT (green triangles), has an even higher steady-state number. Errorbars represent the standard deviation over five repetitions. The dashed and solid lines represent fits considering only one-body loss and accounting for two-body loss, respectively.
    (b) The atom number decreases over time in the MOT after the atomic flux is switched off. Compared with the one-body loss model (dashed lines), the fit that accounts for two-body loss (solid lines) agrees better with the data. The inset presents a complete log-log plot.
\label{fig:loadingrate}}
\end{figure} 

Figure~\ref{fig:loadingrate}(a) shows the loading in the single-color and two-color MOTs. Two-color MOT has a much faster loading rate than the green MOT, while compared with the blue MOT, it has a three-fold increase in steady-state atom number due to shielding enhancement. We fitted the data using models that account for one-body loss (dashed lines) and two-body loss (solid lines). The one-body loss model is $N(t) = \eta/\gamma\times (1-\exp(-\gamma t))$ with $\gamma$ the one-body loss, where the steady-state atom number can be approximated as $N=\eta/\gamma$. 
The two-body loss model, derived from Eq.~\ref{rate equation}, is given by~\cite{caires2004intensity}

\begin{align}
    N(t) = \frac{\sqrt{4\eta\beta+\alpha^2}}{2\beta}\left[\tanh \left(\frac{1}{2}\sqrt{4\eta\beta+\alpha^2}t-\right.\right.
    \notag
    \\ \left.\left.\frac{1}{2}\ln{ \left(-\frac{\alpha-\sqrt{4\eta\beta+\alpha^2}}{\alpha+\sqrt{4\eta\beta+\alpha^2}}\right)} \right)-\frac{\alpha}{\sqrt{4\eta\beta+\alpha^2}}\right]. \label{loading}
\end{align}

\begin{table}[htbp]
\centering
    
\begin{tabular}{lcccc}
\toprule[1pt]
&Blue  &  Green  &  Two-color  \\
   \midrule[1pt]
\makecell[l]{Steady-state number $N$ ($\times10^7$)}&4.1& 8.1 & 11.5\\
\makecell[l]{Loading rate $\eta$ ($\times10^7 \,\text{s}^{-1}$)}&  9.8 &0.10 & 8.4 \\
Loss rate $\gamma$ ($\text{s}^{-1}$)& 2.8& 0.012 & 0.79\\
\makecell[l]{One-body loss $\alpha_{\text{load}}$ ($\text{s}^{-1}$)} &1.3&0.012 & 0.53\\
\makecell[l]{One-body loss $\alpha_{\text{hold}}$ ($\text{s}^{-1}$)}&0.54 & 0.012 &0.21\\
\makecell[l]{Two-body loss $\beta_{\text{load}} $($\times10^{-9}\,\text{s}^{-1}$)}&27   & $\simeq0$ & 1.7\\
\makecell[l]{Two-body loss $\beta_{\text{hold}}$ ($\times10^{-9}\,\text{s}^{-1}$)}&44   & $\simeq0$ &1.0\\
\makecell[l]{$\alpha_{\text{load}}+\beta_\text{load}N$ ($\text{s}^{-1}$)} &2.4&0.012 & 0.73\\
\makecell[l]{$\alpha_{\text{hold}}+\beta_\text{hold}N$ ($\text{s}^{-1}$)} &2.3&0.012 & 0.32\\
\bottomrule[1pt]
\end{tabular}
\caption{\textbf{Loading and loss rates of different MOTs.} Numbers are extracted from Fig.~\ref{fig:loadingrate}. The loading $\eta$ and loss rate $\gamma$ are extracted using the one-body loss model. The one-body loss $\alpha$ and two-body loss $\beta$ are extracted using the two-body loss model from both MOT loading and holding measurements. The steady-state number $N$ is consistent between the one-body and two-body loss models.}
\label{tab:loadlossrate}
\end{table}

We further measure the MOT decay by switching off the atomic flux and holding atoms in MOTs in Fig.~\ref{fig:loadingrate}(b). Similarly, dash lines represent only the one-body loss and the solid lines count for the two body loss, which can be described by Eq.~\ref{rate equation} with $\eta = 0$ and the initial atom number $N_0$
\begin{equation}
    N(t) = \frac{\alpha N_0}{(\alpha+\beta N_0) e^{\alpha t}-\beta N_0}.
\end{equation}

One-body and two-body losses are extracted from loading and holding curves in Fig.~\ref{fig:loadingrate}, and summarized in Table~\ref{tab:loadlossrate}. The two-color MOT not only takes advantage of the fast loading rate of the singlet transition but also substantially eliminates one-body loss, leading to an increase of the MOT atom. Compared with the green MOT, the one-body loss rate of the blue MOT is nearly two orders larger than the green one, resulting in the limit of the MOT lifetime of hundreds ms~\cite{honda1999magneto}. However, the initial atom loading of the blue MOT is two order faster than the green MOT.

Using the experimental parameters and the population equation  $s/(2+s)$, populations of states are $N_{^1\text{P}_1} : N_{^1\text{S}_0} : N_{^3\text{P}_1} = 0.07:1.0:0.99$. The population fractions in the $^1\text{P}_1$ state are $P=N_\text{b}/(N_\text{s}+N_\text{b})= 0.065$ for the blue and $P^{\prime}=N_\text{b}/(N_\text{s}+N_\text{b}+N_\text{g}) = 0.034$ for the two-color MOT. By neglecting $\gamma_c\le0.012\,\text{s}^{-1}$ indicated by the loss rate in the green MOT, the change in the one-body loss rate due to the shielding effect can be estimated by the difference between the loss rates of the blue and two-color MOTs in the holding measurements, $\Delta \alpha 
= 0.33\,\text{s}^{-1}$, in Table~\ref{tab:loadlossrate}, which agrees with the calculated value of $(1-P^{\prime}/P)\alpha_{\text{hold}}=0.26\,\text{s}^{-1}$ based on the loss rate of the blue MOT. The results suggest that the shielding effect arises from a reduction in the one-body loss rate due to redistribution into the intercombination state $^3\text{P}_1$, which acts as a trapped reservoir in the presence of the blue MOT.

Two-body losses influence higher densities, particularly during the saturated loading stage and the initial holding stage in Fig.~\ref{fig:loadingrate}. The two-body losses also depend on the atomic transitions. The two-body losses are almost zero for both loading and holding measurements of the green MOT, whereas they are significant in the blue MOT, shown in Table~\ref{tab:loadlossrate}. The two-color MOT falls in between, due to the intercombination state. As the density further increases, the atomic distribution becomes more uniform rather than Gaussian, and the loss scales as $\propto N$ rather than $\propto N^2$~\cite{weiner1999experiments}.
Interestingly, the sum of the one-body loss rate and the product of the steady-state atom number with the two-body loss coefficient $\alpha+\beta N$ is similar between the loading and holding measurements, consistent with the loss $\gamma$ in the simple one-body loss model.

\section{Conclusion}
We have demonstrated a two-color MOT for fast loading and efficient cooling of ytterbium atoms. We systematically study the mechanisms of the two-color MOT by probing the loading and loss rates for the singlet transition ($^1\text{S}_0\rightarrow {}^1\text{P}_1$), triplet transition ($^1\text{S}_0\rightarrow {}^3\text{P}_1$), and two-color MOTs, and highlight that it benefits from both broad singlet and narrow intercombination transitions. The singlet transition provides rapid loading, while the intercombination transition not only enables further cooling but also shields atoms from losses. We systematically investigate both one-body and two-body losses during atom loading and holding measurements, and experimentally demonstrate that two-body losses start to play a role at high atomic densities. The combination of Zeeman slowing and two-color MOT demonstrated in this work is general and can be extended to other isotopes of ytterbium, as well as to other alkaline-earth(-like) atoms with narrow-line transitions, such as strontium, dysprosium, and erbium, providing a rapid and efficient method for preparing large cold-atom samples for quantum simulation and computing.

\section{Acknowledgments} 
This work was supported by the National Natural Science Foundation of China (12374242), the Innovation Program for Quantum Science and Technology (2024ZD0301800) and the Beijing Natural Science Foundation (Z240007).

\section{Data Availability}
The data presented in this work are available on the Peking University Open Research Data Platform \cite{dataURL}.

\bibliography{MOT.bib}

\end{document}